\global\def\draftcontrol{0}

   \def\versionno{ cpone  }

\catcode`\@=11

\expandafter\ifx\csname draftcontrol\endcsname\relax\global\def\draftcontrol{0}
\fi

{\count255=\time\divide\count255 by 60
\xdef\hourmin{\number\count255}
\multiply\count255 by-60\advance\count255 by\time
\xdef\hourmin{\hourmin:\ifnum\count255<10 0\fi\the\count255}}
\def\draftdate{\number\month/\number\day/\number\year\ \ \ \hourmin }

\newcommand\makepapertitle{\par
  \begingroup
    \renewcommand\thefootnote{\@fnsymbol\c@footnote}%
    \def\@makefnmark{\rlap{\@textsuperscript{\normalfont\@thefnmark}}}%
    \long\def\@makefntext##1{\parindent 1em\noindent
            \hb@xt@1.8em{%
                \hss\@textsuperscript{\normalfont\@thefnmark}}##1}%
     \newpage
     \global\@topnum\z@   
     \@makepapertitle
     \thispagestyle{empty}\@thanks
  \endgroup
  \setcounter{footnote}{0}%
  \global\let\thanks\relax
  \global\let\makepapertitle\relax
  \global\let\@makepapertitle\relax
  \global\let\@thanks\@empty
  \global\let\@author\@empty
  \global\let\@date\@empty
  \global\let\@title\@empty
  \global\let\title\relax
  \global\let\author\relax
  \global\let\date\relax
  \global\let\and\relax
  \def\version{\let\version\@version\@gobble}
}
\def\@makepapertitle{%
  \newpage
   \ifnum\draftcontrol=1 {}
   \version\versionno
   \vskip 3em%
   \else
   \hfill\hbox to 3cm {\parbox{4cm}{\@pubnum}\hss}%
   \vskip 3em%
   \fi
   \begin{center}%
   \let \footnote \thanks
     {\LARGE {\@title}}%
     \vskip 1.5em%
     {\normalsize
       \lineskip .5em%
       \begin{tabular}[t]{c}%
         \@author
       \end{tabular}\par}%
     \vskip 1.5em%
     {\@bstract}%
     \end{center}%
     \vskip 1.5em
     \@date%
   \par
}

\gdef\@pubnum{}
\def\pubnum#1{%
  \gdef\@pubnum{#1}}

\gdef\@bstract{}
\def\Abstract#1{%
  \gdef\@bstract{%
   \parbox{\textwidth-0pc}{%
   \centerline{\bf Abstract}\penalty1000
   \noindent
   \renewcommand\baselinestretch{1.0}
   {#1}}}
}

\def\ps@paper{\let\@mkboth\@gobbletwo%
     \ifnum\draftcontrol=1
        \def\@oddfoot{\hbox to \textwidth{\tiny \versionno \hfil\tiny\draftdate}%
        \hskip -\textwidth \hbox to \textwidth{\hfil\rm\thepage\hfil}}%
     \else\def\@oddfoot{\hbox to \textwidth{\hfil\rm\thepage\hfil}}
     \fi
     \let\@evenfoot\@oddfoot
}

\newenvironment{acknowledgments}{%
\vskip 3.25ex
\noindent {\bf Acknowledgments}
}

\def\@version#1{\ifnum\draftcontrol=1
\typeout{}\typeout{#1}\typeout{}
\vskip3mm\centerline{\hbox{\fbox{\normalsize{\tt DRAFT -- #1 -- }
                   {\draftdate}}}}\vskip3mm
\fi}
\let\version\@version
\long\def\eqlabel#1{\ifnum\draftcontrol=1
                    \tag@false  
                    \tag*{(\theequation) \hbox to -0.2cm{\hspace{0cm}\small{#1}\hss}}
                    \refstepcounter{equation}
                    \edef\@currentlabel{\theequation}
                    \ltx@label{#1}          
                    \else
                    \label{#1}
                    \fi
                    }
\let\st@bibitem\@bibitem
\let\st@lbibitem\@lbibitem
\ifnum\draftcontrol=1
  \def\@bibitem#1{%
    \st@bibitem{#1}\a@@label{#1}\ignorespaces}
  \def\@lbibitem[#1]#2{%
    \st@lbibitem[#1]{#2}\a@@label{#2}\ignorespaces}
  \def\a@@label#1{%
    \gdef\a@lab{\smash{\normalfont\small#1}}
    \ifvmode
      \if@inlabel
        \global\setbox\@labels\hbox{%
          \llap{\a@lab\let\a@lab\relax
                \kern\@totalleftmargin\kern\marginparsep}%
          \box\@labels}%
      \fi
    \fi}
\fi

\documentclass[12pt,letterpaper]{article}

\usepackage{amsmath,amssymb,array,calc,rotating,epsfig,psfrag}
\usepackage[nosort]{cite}

\ifnum\draftcontrol=1
\fi

\renewcommand\baselinestretch{1.25}
\renewcommand\section{\@startsection {section}{1}{\z@}%
                                   {-3.5ex \@plus -1ex \@minus -.2ex}%
                                   {2.3ex \@plus.2ex}%
                                   {\normalfont\large\bfseries}}
\renewcommand\subsection{\@startsection{subsection}{2}{\z@}%
                                   {-3.25ex\@plus -1ex \@minus -.2ex}%
                                   {1.5ex \@plus .2ex}%
                                   {\normalfont\normalsize\bfseries}}
\renewcommand\subsubsection{\@startsection{subsubsection}{3}{\z@}%
                                   {-3.25ex\@plus -1ex \@minus -.2ex}%
                                   {1.5ex \@plus .2ex}%
                                   {\normalfont\normalsize\it}}
\renewcommand\paragraph{\@startsection{paragraph}{4}{\z@}%
                                   {-3.25ex\@plus -1ex \@minus -.2ex}%
                                   {1.5ex \@plus .2ex}%
                                   {\normalfont\normalsize\bf}}
\renewcommand\subparagraph{\@startsection{subparagraph}{5}{\z@}%
                                   {-1.25ex\@plus -1ex \@minus -.2ex}%
                                   {0ex \@plus .2ex}%
                                   {\normalfont\normalsize\it}}

\def\ie{{\it i.e.}}
\def\eg{{\it e.g.}}

\def\revise#1       {\raisebox{-0em}{\rule{3pt}{1em}}%
                     \marginpar{\raisebox{.5em}{\vrule width3pt\
                     \vrule width0pt height 0pt depth0.5em
                     \hbox to 0cm{\hspace{0cm}{%
                     \parbox[t]{4em}{\raggedright\footnotesize{#1}}}\hss}}}}

\def\calm         {{\cal M}}

\def\calo         {{\cal O}}

\def\complex      {{\mathbb C}}

\def\projective   {{\mathbb P}}

\def\reals        {{\mathbb R}}
\def\zet          {{\mathbb Z}}

\def\be{\begin{equation}}
\def\ee{\end{equation}}
\def\beq{\begin{equation}}
\def\eeq{\end{equation}}
\def\bea{\begin{eqnarray}}
\def\eea{\end{eqnarray}}

\def\del          {\partial}

\def\ee           {{\rm e}}

\def\id           {{\bf 1}}

\def\sqr#1#2{{\vcenter{\vbox{\hrule height.#2pt
 \hbox{\vrule width.#2pt height#1pt \kern#1pt
 \vrule width.#2pt}\hrule height.#2pt}}}}

\def\CP{\complex\projective}
\def\RP{\reals\projective}
\def\llangle{\langle\!\langle}
\def\rrangle{\rangle\!\rangle}

\catcode`\@=12

\begin{document}


\begin{titlepage}
\hfill
\vbox{
    \halign{#\hfil         \cr
           arxiv:1106.2967  \cr
           } 
      }  
\vspace*{20mm}

\begin{center}

\hbox to \textwidth{\Large {\bf Open/Closed Topological $\CP^1$ Sigma Model Revisited}\hss}

\vspace*{15mm}

{\sl Shmuel Elitzur $^{a}$, Yaron Oz  $^{b}$,
Eliezer Rabinovici $^{a}$ and Johannes Walcher   $^{c}$}
 \\[0.4cm]
{\it $^{a}$ Racah Institute of Physics, The Hebrew University of Jerusalem, 91904, Israel\\
\it $^{b}$ Raymond and Beverly Sackler School of Physics and Astronomy \\
\it Tel-Aviv University, Ramat-Aviv 69978, Israel \\
\it $^{c}$ Department of Physics, CERN - Theory Division,\\
\it  CH-1211 Geneva 23, Switzerland}

\end{center}

\vspace*{8mm}

\begin{abstract}

We consider the topological sigma-model on Riemann surfaces with
genus $g$ and $h$ holes, and target space $\CP^1\cong S^2$. We
calculate the correlation functions of bulk and boundary operators,
and study the symmetries of the model and its most general deformation.
We study the open/closed topological field theory (TFT) correspondence by summing up the boundaries.
We argue that this summation can be understood as a renormalization of the
closed TFT. We couple the model to topological gravity and derive constitutive
relations between the correlation functions of bulk and boundary operators.

\end{abstract}
\vfill

November 2011

\end{titlepage}

\setcounter{footnote}{0}

\section{Introduction}

In the first quantized string theory one often considers a string moving
in a given geometrical background. One then obtains S-matrix elements
by adding up contributions of worldsheet calculations with different
worldsheet genera. In order to obtain a target space picture, one suggests
a certain effective Lagrangian defined on the world volume, which was the target
space in the worldsheet formulation \cite{Green:1987cw}. This candidate is validated by comparing
the S-matrix elements it produces to those obtained from the worldsheet
procedure. This straightforward procedure does not address various questions such as
the uniqueness of the effective lagrangian and its worldvolume and topology.
One is actually familiar with symmetries such as T-duality and dynamical principles such as
holography, which reflect ambiguities of the effective Lagrangian.
In this work, we study this issue in a very simple setup, which does allow one
to obtain the exact worldsheet results. This is the case of topological theories
of matter.

Topological field theories (TFTs) provide a simple framework to
study open/closed duality properties of string theory. One class of
TFTs are the topological $\sigma$-models \cite{witten}. In order to construct
these models we start with a $(2,2)$ supersymmetric non-linear
$\sigma$-model in two-dimensions. This is a theory of maps $\Phi$
from a two-dimensional worldsheet $\Sigma$ to a target space $X$,
which is a Kahler manifold. The two-dimensional $(2,2)$ theory has a
$U(1)\times U(1)$ R-symmetry. One can twist the theory by adding
to the stress tensor of the theory a derivative a $U(1)$ R-current.
There are two ways to do that, i.e. twisting with the vector symmetry $U(1)_V$
or twisting with the axial symmetry $U(1)_A$. The first leads to the
topological A-model, while the second to the topological B-model \cite{Witten:1991zz}.
Due to the axial anomaly, the B-model is well defined only when the target
space is a Calabi-Yau manifold. We will consider the topological A-model.

After the twisting, the supersymmetry transformation becomes a
transformation under a nilpotent operator $Q$. The action becomes
\beq S \sim \int_{\Sigma}d^2z\{Q,\Lambda\} + t
\int_{\Sigma}\Phi^*(K) \ , \eeq where $\Phi^*(K)$ is the pullback to the
worldsheet of the target space Kahler two-form. We will normalize
\beq \int_{\Sigma}\Phi^*(K) = n \ ,\eeq where $n$ is an integer, the
degree of the instanton. The path integral of the theory localizes
on holomorphic maps, and the correlation functions of the model
depend only on the cohomology class of the Kahler form $K$.

In this paper we will consider the topological sigma-model on Riemann surfaces
with genus $g$ and $h$ holes, and target space $\CP^1\cong S^2$.  We calculate
the correlations function of bulk and boundary operators, study their symmetries
and the open/closed TFT correspondence.
The open/closed topological $\CP^1$ model has been studied several times in the past, in
the context of open/closed topological string correspondence and otherwise. (The earliest
study we are aware of is \cite{horava}.)

The paper is organized as follows.
 We will begin
in section 2 by reviewing the elementary TFT correlation functions of the model,
following \cite{witten,hori}. In section 3, we present our solution of the model at
higher worldsheet topologies. In section 4, we analyze the duality properties of
our results. In section 5 we couple the $\CP^1$ model to topological gravity and
present a few steps towards a complete study of the model. In particular, we
obtain the constitutive relations of the disk amplitude on the large (closed string) phase
space.
Section 6 is devoted to a discussion.
Part of the results on matter TFTs in this paper have been presented in \cite{Elitzur:2009zz}.

\section{The computational scheme}

We will denote by $\langle O_1...O_n \rangle_{g,h}$ the correlation
function of the operators $O_1,...O_n$ on a Riemann surface with
genus $g$ and $h$ boundaries. In this section we will outline
the computational scheme that we will use in order to calculate
these correlation functions.

\subsection{The sphere}

Consider first the correlators on the sphere with no boundaries.
There are two operators of the topological $\CP^1$ $\sigma$-model,
the identity operator, $\id$, and the operator corresponding to the
second cohomology class of the sphere, which we shall denote by $H$.
It is represented by a $\delta$-function two-form on $\CP^1\cong
S^2$. On the worldsheet $H$ is a zero-form. We have \cite{witten}
\begin{equation}
\eqlabel{bulktree}
\begin{split}
\langle \id \rangle_{0,0} &=0 \\
\langle H \rangle_{0,0} &= 1  \\
\langle H^2 \rangle_{0,0} &=0 \\
\langle H^3 \rangle_{0,0} &= \beta \ ,
\end{split}
\end{equation}
where $\beta = e^{-t}$ comes from the classical action of the
$\CP^1$ $\sigma$-model and is the contribution of one-instanton,
i.e. a degree one holomorphic map from the worldsheet to the $\CP^1$
target space. The
one-point correlator of $H$ is constant since it gets contribution
only from the constant map: we map the worldsheet two-sphere to the
target space two-sphere with the point where $H$ is inserted on the
worldsheet being mapped to a given point in the target space. The
three-point correlator of $H$ gets contribution from a degree one
holomorphic map: we map the three insertion points to three given
points on the target space.
From \eqref{bulktree}, one derives the non-trivial ring relation
(OPE)
\begin{equation}
H^2 = \beta \id \ .
\end{equation}

\subsection{The disk}

We now want to consider the $\CP^1$ model with branes included in
the background. As shown in \cite{hori}, there are two
possible branes that preserve topological invariance. Geometrically,
both of them correspond to the equator of $\CP^1$, viewed as
the two-sphere. The two branes are distinguished by the value
$\epsilon=\pm 1$ of a Wilson line.

We consider correlators on the disc, with boundary condition
corresponding to one of the two branes. Both branes support, in
addition to the identity (which we continue to denote by $\id$), a
boundary operator corresponding to the first cohomology class of the
equator circle \cite{hori}, and we will denote this operator by $E$. It is
represented by a $\delta$-function one-form on the equator, and is a
zero-form on the worldsheet. It is shown in \cite{hori} that
\begin{equation}
\eqlabel{boundarytree}
\begin{split}
\langle \id \rangle_{0,1} &= 0 \\
\langle E\rangle_{0,1} &= 1 \\
\langle H\rangle_{0,1} &= 0 \\
\langle E^2\rangle_{0,1} &= 0 \\
\langle E^3\rangle_{0,1} &= \epsilon\beta^{1/2} \\
\langle E H \rangle_{0,1} &= \epsilon\beta^{1/2} \ .
\end{split}
\end{equation}
Here, it is understood that $E$ will be inserted on the boundary,
while $H$ is inserted in the bulk of the disc. The one-point
correlator of $E$ on the disk is constant since it receives
contribution only from the constant map: we map the disk worldsheet
to the target space $S^2$, such that the boundary of the disk is
mapped to the equator and the insertion point on the boundary of the
disk is mapped to a given point on the equator. This holomorphic map
is the constant map. The three-point correlator of $E$ receives
contribution from the disk one-instanton (or half-instanton in the
closed string sense), which is a degree one map from the disk to
$S^2$, where the boundary of the disk is mapped to the equator. The
three insertion points on the boundary of the disk are mapped to
three points on the equator.

The equations \eqref{boundarytree} imply the non-trivial relation on
the boundary
\begin{equation}
E^2 = \epsilon \beta^{1/2} \ ,
\end{equation}
as well as the important relation
\begin{equation}
\eqlabel{important}
H = E^2 = \epsilon\beta^{1/2}\id
\end{equation}
between the bulk field $H$ and the boundary field $E^2$.
In other words, when computing a correlator on a Riemann surface
with boundary, an insertion of $H$ in the bulk is equivalent to
inserting $E^2$ on the boundary (or equivalently, inserting
$\epsilon\beta^{1/2}\id$, in the bulk or on the boundary).

It is also important to note that there are no boundary condition
changing operators between $\epsilon=+1$ and $\epsilon=-1$.

\subsection{Axial R-charge}

We assign axial R-charges to the operators \beq R[H] = 2,~~~~ R[E] =
1 \ . \eeq In general, an amplitude $\langle H^n E^m\rangle_{g,h}$ will
be proportional to $\beta^{k}$ if $k$ is an integer (without
boundaries) or half-integer (with boundaries) satisfying
\begin{equation}
\eqlabel{satisfying}
\sum_i R_i = 2 n + m = 4 k + 2 - 2 g - h \ .
\end{equation}
If there is no such $k$ the amplitude vanishes.

Note that a $(g,h)$ amplitude will vanish if both boundary
conditions appear at the boundaries of the surface. This is because
there are no boundary condition changing operators. Therefore, we
will choose $\epsilon$ equal for all boundaries and fix it.
Moreover, the amplitude will vanish unless $E$ is inserted an odd
number of times on each boundary.

\subsection{Handle and boundary states}

An important idea of \cite{witten}, picked up in \cite{efr}, is that the
TFT correlation functions on Riemann surfaces of higher genus can be computed
as correlation functions on the sphere with some additional insertions. This
idea can be straightforwardly generalized to the present situation with
boundaries.

Thus, we introduce a handle operator $W$, with defining property.
\begin{equation}
\langle \calo \rangle_{g,h} =
\langle W \calo \rangle_{g-1,h} \ ,
\end{equation}
\ie, it relates correlators on surfaces of different topology.
One can compute $W$ at $g=1$ by degenerating the torus
into a sphere. Here, the relations
\begin{equation}
\begin{split}
\langle W \rangle_{0,0} &= \langle \id\rangle_{1,0}
= 2 \langle H\rangle_{0,0} = 2 \\
\langle W H \rangle_{0,0} & = \langle H \rangle_{1,0}
= 2\langle H^2\rangle_{0,0} = 0 \ ,
\end{split}
\end{equation}
imply
\begin{equation}
W = 2 H \ .
\end{equation}

When considering surfaces with boundaries, we have to fix
boundary conditions $a_1, a_2, \ldots, a_h$ on each of
them. Moreover, we have to allow for a dependence of the
boundary state on the boundary insertions. We will
then label the boundary states as $V_{a,\theta}$
to indicate dependence on the boundary condition and
the boundary insertion. It satisfies
\begin{equation}
\langle \calo \theta_1 \cdots \theta_h \rangle_{g,h} = \langle \calo
\theta_1\cdots \theta_{h-1} V_{a_h,\theta_h} \rangle_{g,h-1} \ .
\label{boundaryr}
\end{equation}
It should be understood that the operator $\theta_i$ is inserted
on the $i$-th boundary on the left and on the right hand side
of the equation.

The boundary states can be computed on the disc. From
\eqref{boundarytree}, we learn
\begin{equation}
\begin{split}
V_{\epsilon,\id} &= 0 \\
V_{\epsilon,E} &= \epsilon\beta^{1/2}\id + H \ .
\end{split}
\label{Ve}
\end{equation}

\subsection{Frobenius algebra}

Field theories can be axiomatized by the algebra structure provided
by their operators. For a TFT on closed Riemann surfaces, the
relevant structure is that of a Frobenius algebra (we consider the quantum algebra
deformed by the worldsheet instantons). For the $\CP^1$
model, the Frobenius algebra has a basis of idempotents, which are
given by
\begin{equation}
H_{\pm}= \frac12(\id\pm\beta^{-1/2} H) \ ,
\end{equation}
and satisfy the algebra
\begin{equation}
\eqlabel{algebra}
H_{+}^2 = H_+ \qquad H_{-}^2 = H_- \qquad
H_+ H_- = 0 \ .
\end{equation}
The trace on the algebra is given by
\begin{equation}
\eqlabel{eta}
\langle H_\pm \rangle_{0,0} = \eta_\pm \equiv \pm\frac 1{2 \beta^{1/2}} \ .
\end{equation}
The handle operator is
\begin{equation}
W= \frac{H_+}{\eta_+} + \frac{H_-}{\eta_-} =
2\beta^{1/2} (H_+-H_-) \ .
\end{equation}

Abstractly, branes should correspond to modules over the algebra
of bulk operators, in other words, to (irreducible) representations
of this algebra. Indeed, using \eqref{important}, we learn that,
in presence of the boundary $\epsilon$,
\begin{equation}
H_\epsilon = \id\qquad\qquad H_{-\epsilon} = 0 \ ,
\end{equation}
which are indeed the possible representations of \eqref{algebra}.

Note also that
\begin{equation}
V_{\epsilon,E}= 2\epsilon\beta^{1/2} H_\epsilon =
\frac{H_\epsilon}{\eta_\epsilon} \ . \label{V}
\end{equation}

\section{Solving the $\CP^1$ model}

In this section we will compute the exact correlation functions of
the $\CP^1$ model. Here, by ``exact'', we mean that we will sum over
worldsheet topologies, but without coupling to topological gravity.

\subsection{Summing over genera}

We weight a closed Riemann surface of genus  $g$ by a factor
$\lambda_c^{2g-2}$, where $\lambda_c$ is the  closed string
coupling. Without boundaries we have \cite{witten,efr}
\begin{equation}
\begin{split}
\llangle \id \rrangle &\equiv \sum_{g=0}^\infty \lambda^{2g-2}_c
\langle \id \rangle_{g,0} = \sum_{g=0}^\infty \lambda^{2g-2}_c
\langle W^g\rangle_{0,0} =
\sum_{g\text{ odd}} \lambda_c^{2g-2}\langle (2H)^g\rangle_{0,0} \\
& =\sum_{n=0}^\infty \lambda^{4n}_c 2^{2n+1} \beta^n
= \frac{2}{1- 4 \lambda^4_c \beta} \ ,
\end{split}
\end{equation}
and
\begin{eqnarray}
\llangle H^{2s} \rrangle
&=&\sum_{n=0}^\infty \lambda^{4n}_c 2^{2n+1} \beta^{n+s}
= \frac{2\beta^s}{1- 4 \lambda^4_c \beta} \ , \nonumber\\
\llangle H^{2s+1} \rrangle
&=&\sum_{n=0}^\infty \lambda^{4n-2}_c 2^{2n} \beta^{n+s}
= \frac{\lambda_c^{-2}\beta^s}{1- 4 \lambda^4_c \beta} \ .
\end{eqnarray}
In the idempotent basis \eqref{algebra}, this can be written as
\begin{equation}
\begin{split}
\llangle  H_\epsilon  \rrangle &= \sum_{g=0}^\infty \lambda^{2g-2}_c
\langle H_\epsilon W^g\rangle_{0,0}
= \sum_{g=0}^\infty  \lambda^{2g-2}_c \eta_\epsilon^{1-g}
= \frac{\lambda^{-2}_c \eta_\epsilon}{1-\lambda^{2}_c\eta_\epsilon^{-1}} \\
&= \frac{1}{\frac{\lambda^{2}_c}{\eta_\epsilon}
\bigl(1-\frac{\lambda^{2}_c}{\eta_\epsilon}\bigr)} \ .
\end{split}
\end{equation}
This result is seemingly invariant under \cite{efr}
\begin{equation}
\frac{\lambda^{2}_c}{\eta_\pm} \leftrightarrow 1-\frac{\lambda^{2}_c}{\eta_\pm} \ .
\label{symmetry}
\end{equation}

Note, however, that according to \eqref{eta}, $\eta_+=-\eta_-$ and the transformations
\eqref{symmetry} are not mutually compatible. Thus, only correlators of one
type, say $\llangle H_+  \rrangle$ are invariant and, in particular,
$\llangle \id \rrangle$ is not invariant as can be easily checked.
One can, however, deform the theory as to make it completely
invariant. On general grounds, one expects the theory to depend on as many
parameters as there are operators in the theory. As explained in \cite{efr},
these parameters are most easily encoded in the sphere one-point functions of
the basis of idempotents, while keeping fixed the rest of the OPE.
In our case, we write
\begin{equation}
\eqlabel{ourcase}
\langle H_+\rangle_{0,0} = \tilde\eta_+  \,,
\qquad
\langle H_- \rangle_{0,0} = \tilde\eta_-
\end{equation}
and we may in general treat $\tilde\eta_-$ as independent from $\tilde\eta_+$. We recover
the standard $\CP^1$ model on the subspace $\tilde\eta_-=-\tilde\eta_+$. Note that this
deformation may or may not be realized in the standard BRST procedure, and may not survive
coupling to topological gravity.

Now repeating the above computation we have
\begin{equation}
\llangle  H_\epsilon  \rrangle =
\frac{1}{\frac{\lambda^{2}_c}{\tilde\eta_\epsilon}
\bigl(1-\frac{\lambda^{2}_c}{\tilde\eta_\epsilon}\bigr)} \ .
\end{equation}
Since now $\tilde\eta_+$ and $\tilde\eta_-$ are independent
parameters, we have an exact symmetry of the theory generated by

\let\varepsilon\epsilon
\begin{equation}
\frac{\lambda^{2}_c}{\tilde\eta_\varepsilon}\rightarrow
\frac{\lambda^{2}_c}{\tilde\eta_\varepsilon},~~~~
\frac{\lambda^{2}_c}{\tilde\eta_{-\varepsilon}}
\leftrightarrow 1-\frac{\lambda^{2}_c}{\tilde\eta_{-\varepsilon}} \ .
\label{symmetry1}
\end{equation}

Note that although we seem to have introduced three parameters
$(\tilde\eta_+,\tilde\eta_-,\lambda)$,
the correlation functions on closed Riemann surfaces depend only
on two parameters, which are the combinations
$(\frac{\lambda^{2}_c}{\tilde\eta_\varepsilon},
\varepsilon = \pm)$.

\subsection{The annulus}

\begin{figure}[htb]
\psfrag{t1}{$\theta_1$}
\psfrag{t2}{$\theta_2$}
\psfrag{H}{$H$}
\psfrag{1}{$\id$}
\psfrag{plus}{$+$}
\psfrag{E}{$E$}
\begin{center}
\epsfig{file=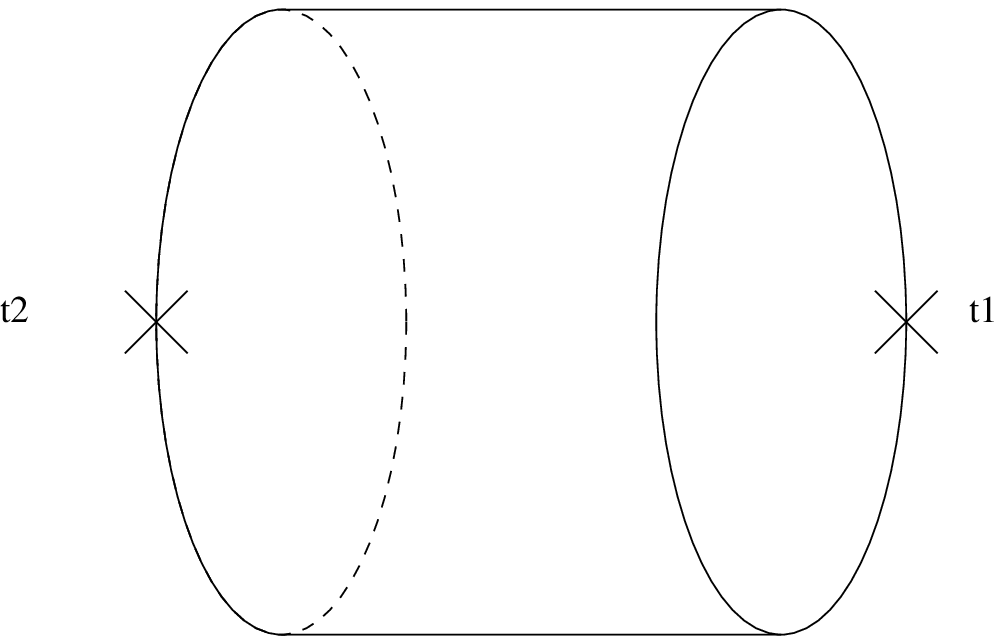,height=2cm} \\[.4cm]
\epsfig{file=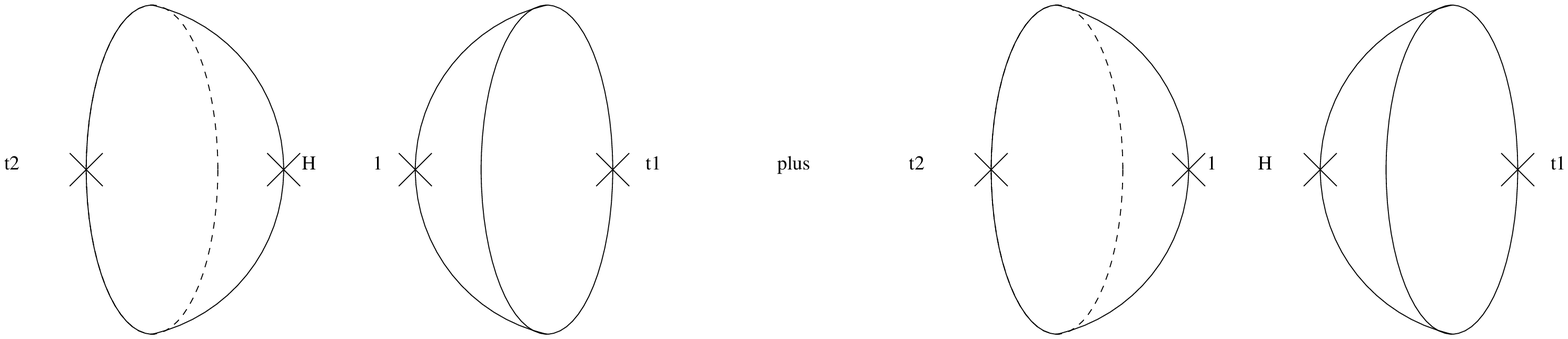,height=2cm} \\[.4cm]
\epsfig{file=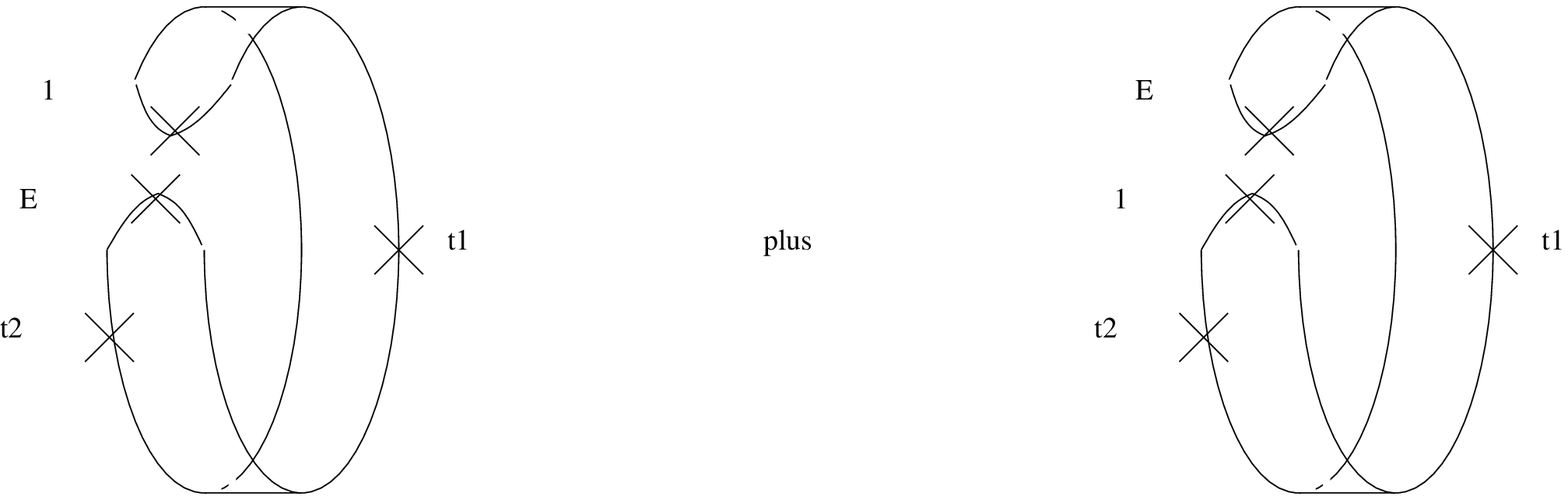,height=2cm}
\end{center}
\caption{Factorizing the annulus}
\label{factor}
\end{figure}

As a warmup for higher genus computations with background D-branes, let us check
the factorization properties of the annulus correlator (see Fig.\ \ref{factor}).
We put equal boundary conditions on the two boundaries. There are then three
amplitudes to consider: $\id$ inserted on both boundaries,
$\id$ on one, $E$ on the other boundary, or $E$ on both
boundaries.

Factorizing via boundary states (middle of Fig.\ \ref{factor}), we
find
\begin{equation}
\eqlabel{annulus}
\begin{split}
\langle \id \id \rangle_{0,2} &= 0 \\
\langle \id E \rangle_{0,2} &= 0 \\
\langle E E \rangle_{0,2} &= 2\epsilon\beta^{1/2} \ .
\end{split}
\end{equation}

We can also factorize as in the bottom of Fig.\ \ref{factor}.
\begin{equation}
\langle \id \id \rangle_{0,2} = \langle\id E \rangle_{0,1} \pm
\langle E \id \rangle_{0,1} \ ,
\end{equation}
where $\pm$ is a sign that appears to be not so well understood
in the general axiomatics of open-closed TFT (see,\eg,
\cite{lazaroiu}). To be consistent, we here need this sign to be
$-$. The second correlator
\begin{equation}
\langle \id E \rangle_{0,2} = \langle\id E^2\rangle_{0,1} \pm
\langle E\id E\rangle_{0,1} = 0
\end{equation}
does not suffer from this ambiguity and is consistent with
\eqref{annulus} in any case. On the other hand,
\begin{equation}
\langle E E\rangle_{0,2} = \langle E^3 \rangle_{0,1} \pm
\langle E^3 \rangle_{0,1}
\end{equation}
requires $\pm=+$, in which case, using \eqref{boundarytree}, we
obtain agreement with \eqref{annulus}. We conclude that the
relative sign between the two terms in the last line of Fig.\
\ref{factor} depends on the boundary insertions.

\subsection{General amplitudes}

We now turn to a computation of exact correlation functions in
the presence of background D-branes. In order to carry out this
computation, we have to supply the combinatorial factors involved
in summing over boundaries, and in distributing boundary insertions
over the various boundaries. To remain flexible, we introduce an
open string coupling constant $\lambda_o$ and weigh a worldsheet
of topology $(g,h)$ by $\lambda_c^{2g-2}\lambda_o^h$. Our aim is
to evaluate
\begin{equation}
\eqlabel{note}
\llangle E^m H^n_\epsilon \rrangle
\sim \sum_{g=0,h=1}^\infty \lambda_c^{2g-2}\lambda_o^h
\langle E^m H^n_\epsilon \rangle_{g,h} \ ,
\eeq
where we have written $\sim$ for the time being since we have not yet
specified the combinatorial factors on the RHS.

Using the above handle and boundary operators, we find
\begin{equation}
\eqlabel{proceed}
\langle E^m H_\epsilon^n \rangle_{g,h} =
\langle E^m H_\epsilon^n W^g\rangle_{0,h} =
 \eta_\epsilon^{-g} \langle E^m H_\epsilon\rangle_{0,h} \ ,
\end{equation}
where we use $H_\epsilon^2=H_\epsilon$ and $W=H_+/\eta_++H_-/\eta_-$, and we assume that
all the boundaries carry label $\epsilon$. (As mentioned before, the amplitudes
otherwise vanish.)

Note that to proceed with \eqref{proceed} we have to know how the boundary insertions
are distributed on the various boundaries. This is not specified in \eqref{note},
which involves a sum over all possible numbers of boundaries. Thus, there is an ambiguity
that will accompany  us for the next few pages. To ensure that \eqref{proceed} is
non-vanishing, we assume $h\equiv m\bmod 2$, split the $m$ factors of $E$ into a group of $h$,
to be put one on
each boundary, and the remaining $s=(m-h)/2$ pairs into $h$ groups of arbitrary size.

Assuming
the $E$'s are initially indistinguishable, this introduces a combinatorial factor of
\begin{equation}
\binom{s+h-1}{h-1}
\end{equation}
so that we get:
\begin{equation}
\eqlabel{sothatweget}
\langle E^m H_\epsilon^n \rangle_{g,h} = \binom{s+h-1}{h-1}
\eta_\epsilon^{-g} \bigl(2\eta_\epsilon\bigr)^{-s} (\eta_\epsilon)^{-h} \eta_\epsilon \ ,
\end{equation}
where we use $E^2 = (2\eta_\epsilon)^{-1}$, the boundary state $V_{\epsilon,E}= H_\epsilon/\eta_\epsilon$
as well as $\langle H_\epsilon\rangle_{0,0}=\eta_\epsilon$.

Now the sum over $h$ in \eqref{note} is restricted to those $h$ with the same parity as $m$.
As above, we write $m=2s+h$, and sum over $s$. This yields
\bea
\llangle E^m H^n_\epsilon\rrangle &=& \sum_{g=0}^\infty \sum_{s=0}^{m/2}
\lambda_c^{2g-2} \lambda_o^{m-2s} \eta_\epsilon^{1-g+s-m}2^{-s} \binom{m-s-1}{m-2s-1} \nonumber \\
&=& \frac{1}{\frac{\lambda_c^2}{\eta_\epsilon}\bigl(1-\frac{\lambda_c^2}{\eta_\epsilon}\bigr)}
\sum_{s=0}^{m/2}\lambda_o^{m-2s} \eta_\epsilon^{s-m} 2^{-s} \binom{m-s-1}{m-2s-1} \ .
\eea

The last sum in this expression is a certain hypergeometric polynomial. In order to check whether
the symmetry (\ref{symmetry1}) remains in the presence of boundaries, it is useful
to compute $\llangle E^m H_{\varepsilon} \rrangle$ in the general deformation \eqref{ourcase}
We get in a straightforward fashion:
\begin{equation}
\llangle E^m H^n_\epsilon\rrangle =
= \frac{1}{\frac{\lambda_c^2}{\tilde\eta_\epsilon}\bigl(1-\frac{\lambda_c^2}{\tilde\eta_\epsilon}\bigr)}
\sum_{s=0}^{m/2}\lambda_o^{m-2s} \tilde\eta_\epsilon^{s-m} 2^{-s} \binom{m-s-1}{m-2s-1}
\eqlabel{exactresult} \ .
\end{equation}
We can rescale the boundary operator by $E \rightarrow
(2\eta_\epsilon)^{\frac{1}{2}} E$. We then see that although we have
introduced four parameters $(\tilde\eta_+,\tilde\eta_-,\lambda_c,\lambda_o)$
for the open plus closed system, the correlation functions in the presence
of boundaries depend only on two parameters
$(\frac{\lambda^{2}_c}{\tilde{\eta_\epsilon}},\frac{\lambda_o^2}{\tilde\eta_{\epsilon}})$
(for fixed choice of $\epsilon$).
Note that in the presence of boundaries both signs of $\epsilon$
are not allowed simultaneously. It is then obvious the
symmetry (\ref{symmetry1}) of the closed TFT is preserved by the
open plus closed system, provided we require that we keep
$\frac{\lambda_o^2}{\tilde\eta_{\epsilon}}$ invariant. This works, whether
or not we relate the open string coupling to the closed coupling,
such as implied by unitarity in string theory. On the other hand, we do not see any
new duality appearing in the open string sector.
It is interesting to note that the sum over boundaries is less singular (a polynomial)
than the sum over genera, which gives a pole at $\frac{\lambda_c^2}{\tilde\eta_\epsilon}=1$.
This may be interpreted as an analogue of the fact that standard closed string (gravity)
perturbation theory is more singular than open string (gauge theory) perturbation theory.
In string theory, this distinction arises because of the properties of the moduli
space of Riemann surfaces after coupling to worldsheet gravity, which we have not done.

\section{Open/Closed duality}

A way to view open/closed string theory duality is that summing up the open string degrees
of freedom results in modifying the closed string background.
Schematically,
\begin{equation}
\sum_{open}{\cal F}_{open+closed}(t_o,t_c) = {\cal F}_{closed}(t'_c) \ .
\end{equation}
Here $t_o$ and $t_c$ denote all the open and closed string moduli,
and $t'_c$ are the modified closed string moduli due to the open
strings back reaction. A natural question is whether we can see such
a duality in our open/closed topological field theory.
We will see that in the absence of worldsheet gravity,
summing up the open string degrees
of freedom results in a renormalization of the closed TFT operator.

\subsection{Generating Functionals}

The generating functional $F_{g,h}(t_{H},t_E)$ for the correlators
of the type $\langle E^m H_{\epsilon}^n \rangle_{g,h}$ is characterized by the
property
\begin{equation}
\frac{\partial^{n+m}F_{g,h}(t_{H},t_E)}{\partial^{n}t_{H_{\epsilon}}
 \partial^{m}t_E}|_{t_{H}=0,t_E=0}
=\langle E^m H_{\epsilon}^n \rangle_{g,h} \ .
\end{equation}
As noted before, $H_{\epsilon}$ is defined with the same $\epsilon$
as the D-brane boundary condition. There are also the observables
$H_{-\epsilon}$. However, correlators $H_{-\epsilon}$ with the other
observables have only the disconnected parts.
Define
\begin{equation}
\eqlabel{openplclosed}
\begin{split}
{\cal F}_{open+closed}(\lambda_c,\lambda_o,t_H,t_E,\eta_\pm) &=
\sum_{g=0}^{\infty}\sum_{h=0}^{\infty}\lambda_{c}^{2g-2}\lambda_{o}^h
F_{g,h}(t_{H},t_E,\eta_\pm)\\ &=
\sum_{m=0}^{\infty}\sum_{n=0}^{\infty} \llangle E^m H_{\epsilon}^n
\rrangle \frac{t_E^m}{m!} \frac{t_{H_{\epsilon}}^n}{n!} \ .
\end{split}
\end{equation}
The generating functional for the correlators of the closed
topological $\sigma$-model reads
\begin{equation}
{\cal F}_{closed}(\lambda_c,t_{H_{\varepsilon}})=
\frac{\frac{\eta_{\epsilon}}{\lambda_c^2}}
{1-\frac{\lambda_c^2}{\eta_{\epsilon}}} \exp[t_{H_{\epsilon}}] \ .
\label{closed}
\end{equation}
In particular, using $\eta_{-\epsilon}= -\eta_{\epsilon}$ for the undeformed
topological $\sigma$-model, we see that
\begin{eqnarray}
\langle {\bf{1}} \rangle_{Exact} = \langle H_{\epsilon}+
H_{-\epsilon}  \rangle_{Exact} =\frac{2}{1-
\Big(\frac{\lambda_c^2}{\eta_{\epsilon}}\Big)^2} \ .
\end{eqnarray}

We can obtain a simple expression for the closed generating
functional after summing up over the boundaries. Recall that we can
replace a boundary with an insertion of the boundary operator $E$ by
using the operator $V_{\epsilon,E}$ as described by equations
(\ref{boundaryr}) and (\ref{Ve}). We need to sum over the boundaries
with any number of (odd) $E's$ inserted on each boundary.
Consider first one boundary. We have
\begin{equation}
\lambda_o \sum_{m=1}^{\infty} \langle H_{\epsilon} E^{2m+1}\rangle_{0,1}\frac{t_E^{2m+1}}{(2m+1)!} =
\langle H_{\epsilon}\rangle_{0,0} \lambda_o \Bigl(\frac{2}{\eta_\epsilon}\Bigr)^{1/2}
\sinh\left[\frac{t_E}{(2\eta_\epsilon)^{1/2}}\right] \ .
\label{onebound}
\end{equation}
Now we need to sum over the number of boundaries, which exponentiates (\ref{onebound}).
This can be achieved in the generating functional by modifying
\beq
\eqlabel{shift}
t_{H_{\epsilon}}
\rightarrow \tilde{t}_{H_{\epsilon}} = t_{H_{\epsilon}} +  \lambda_o
\Bigl(\frac{2}{\eta_\epsilon}\Bigr)^{1/2}
\sinh\left[\frac{t_E}{(2\eta_\epsilon)^{1/2}}\right] \ .
\eeq

Thus, the generating functional for the open plus closed topological
$\sigma$-model (\ref{openplclosed}) is obtained by using the change (\ref{shift}) in the
generating functional of the closed topological $\sigma$-model (\ref{closed}).
We see that summing up the open string degrees of freedom results in a renormalization of the
closed string operator $H_{\epsilon}$ by adding to it an infinite series of the boundary
operator $E$, weighted by $\eta_\epsilon$ and the open string coupling $\lambda_o$.

\subsection{Alternatives}

In this subsection, we explore some alternative combinatorial rules for summing over boundaries,
in view of simplifying open-closed duality. To appreciate these alternatives, one has to
realize that while the set of correlation functions satisfies the axioms of TFTs, we know
of no a priori constraints on how to choose the combinatorial factors in summing over worldsheet 
topologies. This freedom is a consequence of the fact that we do not couple here the TFT
to a worldsheet gravity. A related expectation is that eventually the ambiguities in summing
the holes will be be removed by the uncovering of a new
symmetry that should be maintained by the factors or by some other
consistency argument. In the absence of this
guiding principle we here advocate considering various possibilities mentioning each
time an additional physical input, which would select
that particular choice. We also include one choice, whose sole
present motivation is the interesting result it implies
on open-closed string duality. It displays a property consistent
with our prejudices. The fact that other choices do not lead to that 
result should be kept as a cautionary fact as
long as no new consistency conditions are uncovered.

As before, we assume the background of one D-brane labeled $\epsilon$, and use our formula
\eqref{sothatweget} for the perturbative correlators. We now sum over genera, number of boundaries,
and the number of bulk and boundary insertions. As before, we write $m=2s+h$. This leads to the
full free energy:
\begin{equation}
\begin{split}
F(\lambda_c,\lambda_0,t_{H_\epsilon},t_E)
&= \sum_{g,h,n,s} \frac{t_{H_\epsilon}^n}{n!}\,\frac{t_E^{2s+h}}{(2s+h)!}\, \lambda_c^{2g-2}\,\lambda_o^h
\,\langle E^{2s+h} H_\epsilon^n\rangle_{g,h} \\
&= \sum_{g,h,n,s} \frac{t_{H_\epsilon}^n}{n!}\, \frac{t_E^{2s+h}}{(2s+h)!}\,   \lambda_c^{2g-2}\,\lambda_o^h\,
\frac{(s+h-1)!}{s! (h-1)!} \eta_\epsilon^{1-g-h-s}\\
&= \frac{\eta_\epsilon}{\lambda_c^2}\frac{1}{1-\frac{\lambda_c^2}{\eta_\epsilon}} \, \ee^{t_{H_\epsilon}}
\, \sum_{s,h} \frac{(s+h-1)!}{(2s+h)! s! (h-1)!} \lambda_o^h \eta_\epsilon^{-h-s} t_E^{2s+h} \ .
\end{split}
\end{equation}
The first factor is from geometric sum over genus. The second is the exponential factor
from summing over bulk insertions. The last factor however is from devil's kitchen, and cannot be done
in a closed form. Note however that it does contain $1/(h-1)!$, which might be intuitively expected from
indistinguishability of the boundaries.

We now invoke the right to modify the combinatorial factor involved in summing over boundary insertions. For
example, we might consider replacing $t_E^{m}/m!$ with $t_E^{2s}/(2s)!$ if we decided not to count the
first insertion that goes on each boundary to make the correlator non-zero. A physical way to justify
this modification is to consider D-branes on which we have ``turned on'' the $E$ insertion.
The mathematical advantage is that we are now able to do the sum:
\begin{equation}
\sum_{s,h} \frac{(s+h-1)!}{(2s)! s! (h-1)!} \left(\frac{\lambda_o}{\eta_\epsilon}\right)^h\,
\left(\frac{t_E^2}{\eta_\epsilon}\right)^s =
\frac{y}{1-y} \cosh\sqrt{\frac{x}{1-y}} \ ,
\end{equation}
where $y=\lambda_o/\eta_\epsilon$, and $x=t_E^2/\eta_\epsilon$. So
\begin{equation}
\eqlabel{harmonious}
F(\lambda_c,\lambda_0,t_{H_\epsilon},t_E) = \frac{\ee^{t_{H_\epsilon}}}
{\frac{\lambda_c^2}{\eta_\epsilon}(1-\frac{\lambda_c^2}{\eta_\epsilon})} \;
\frac{\frac{\lambda_o}{\eta_\epsilon}}{(1-\frac{\lambda_o}{\eta_\epsilon})}\cosh\tilde t_E \ ,
\end{equation}
where $\tilde t_E = t_E \sqrt{\frac{\frac{1}{\eta_\epsilon}}{1-\frac{\lambda_o}{\eta_\epsilon}}}$.
This result does not to allow an interpretation as an open-closed string duality, with a closed dual theory being a deformation
of the original TFT.
We are therefore
encouraged to look for other possibilities.

Another possible proposal for the combinatorial factor is to claim that we should only
count boundary insertions in pairs. This would mean putting $t_E^s/s!$ in the sum, which
becomes
\begin{equation}
\sum_{s,h} \frac{(s+h-1)!}{s! s! (h-1)!} x^s y^h = \frac{y}{1-y} \ee^{x/(1-y)} \ ,
\end{equation}
with $y=\lambda_o/\eta_\epsilon$ as before and $x= t_E/\eta_\epsilon$. Then
\begin{equation}
F(\lambda_c,\lambda_o,t_{H_\epsilon},t_E) = \frac{\ee^{t_{H_\epsilon}}}
{\frac{\lambda_c^2}{\eta_\epsilon}(1-\frac{\lambda_c^2}{\eta_\epsilon})} \;
\frac{\frac{\lambda_o}{\eta_\epsilon}}{(1-\frac{\lambda_o}{\eta_\epsilon})}
\ee^{\tilde t_E} \ ,
\end{equation}
with $\tilde t_E = t_E \frac{\frac{1}{\eta_\epsilon}}{1-\frac{\lambda_o}{\eta_\epsilon}}$.
This result is quite similar to \eqref{harmonious}, without a manifest open-closed duality.

By way of answer analysis, one may check that in order to obtain a standard open-closed
duality, we would need the combinatorial factor for summing over boundary insertions to be
\begin{equation}
\eqlabel{wouldneed}
\frac{t_E^s}{(s+h-1)!}
\end{equation}
Then the sum becomes
\begin{equation}
\sum_{s,h}\frac{1}{s!(h-1)!} \left(\frac{\lambda_o}{\eta_\epsilon}\right)^h\;
\left(\frac{t_E}{\eta_\epsilon}\right)^s = \frac{\lambda_o}{\eta_\epsilon}
\ee^{t_E/\eta_\epsilon} \ee^{\lambda_o/\eta_\epsilon} \ .
\end{equation}
In this scheme, the effect of integrating out the
open strings is a shift of the closed string parameter
\begin{equation}
t_{\rm closed} \to t_{\rm closed} + \lambda_o/\eta_\epsilon \ .
\end{equation}
Note, however, that we do not currently have  a physical justification for \eqref{wouldneed}.
Also, the $\ee^{t_E/\eta_\epsilon}$.
 term deserves a better understanding.

\section{Coupling to Topological Gravity}

In this section, we initiate a systematic attempt to couple the open-closed topological
$\CP^1$ model to topological gravity, starting from first principles (\ie, without using
dualities of any sort). To the best of our knowledge, this has not been attempted before,
mostly, it appears, because the notion of open topological gravity is severely
under-developed, and naively mathematically ill-defined. (We believe, however, that
a sensible version exists.) As in the previous sections, we
here take a pragmatic approach, leaving justifications to future work. The key to success
will be to consider only gravitational descendants in the bulk, and not on the boundary.

\subsection{Topological gravity}

When coupling the closed topological $\CP^1$  $\sigma$-model to topological gravity,
we have the following operators:
In the closed sector we have
the puncture operator $P$ that fixes a point in
 the bulk of the worldsheet (creates a puncture),
and the  primary operator $H$.
In addition we
 have the gravitational descendants
$\sigma_{k}(P), \sigma_k (H), k=1,2,...$.

When all couplings are turned off, we
have the following non-vanishing correlators at genus $0$ \cite{witten}
\begin{equation}
\eqlabel{diwieq}
\langle PP H\rangle_{0,0} =  1\,,\qquad \langle H^n \rangle_{0,0}= 1\,,\qquad \text{for all
$n\ge 3$} \ .
\end{equation}
The first equation is the contribution from the degree $0$ sector, and the
second from sector with instanton number $1$. Note the difference to the
correlators without coupling to topological gravity \eqref{bulktree}. Most
succinctly, this change can be seen in the selection rule. Without
coupling to topological gravity, this selection rule is
\begin{equation}
\eqlabel{without}
1 + 2k = n_H \ ,
\end{equation}
where $n_H$ is the number of insertions of $H$, and $k$ is the instanton number
(the degree of the map $\CP^1\to\CP^1$) Coupling to topological gravity modifies
this to
\begin{equation}
\eqlabel{with}
-2 + n_P + n_H + 2k = n_H \ ,
\end{equation}
where $n_P$ is number of insertions of puncture operator. Note that
$n_H$ drops out of \eqref{with}, whereas \eqref{without} does not depend on $n_P$.

Now note that in models such as $\CP^1$, insertions of operators
correspond to ordinary derivatives with respect to the corresponding couplings,
\eg,
\begin{equation}
\del_{t_{0,H}}\langle \cdots\rangle = \langle H \cdots \rangle
\,,\qquad
\del_{t_{0,P}}\langle \cdots\rangle = \langle P \cdots \rangle \ .
\end{equation}
This implies that the relations \eqref{diwieq} can be summarized in the
following generating function (``prepotential'')
\begin{equation}
\eqlabel{prepot}
F^{(0,0)}=\langle \rangle_{0,0} = \frac 12 t_{0,P}^2 t_{0,H} + \ee^{t_{0,H}} \ .
\end{equation}
Equivalently, on the small phase space, we have the correlators
\begin{equation}
\eqlabel{wehave}
\begin{split}
\langle PP\rangle_{0,0} &= t_{0,H} \\
\langle PH \rangle_{0,0} &= t_{0,P} \\
\langle HH \rangle_{0,0} &= \ee^{t_{0,H}}
\end{split}
\end{equation}
To study the large phase space (turn on coupling to descendants), it is a
good idea to rewrite \eqref{wehave} as ``constitutive relations''. Namely,
as emphasized in \cite{diwi}, {\it the functional form of the correlators is
unchanged if we express them as functions of the coordinates}
\begin{equation}
\eqlabel{coordinates}
\begin{split}
u^P &=  \langle PP\rangle_{0,0} = t_{0,H} \\
u^H &= \langle PH \rangle_{0,0} = t_{0,P} \ .
\end{split}
\end{equation}
Using this, the constitutive relation of $\CP^1$ is
\begin{equation}
\eqlabel{constitutive}
\langle HH \rangle_{0,0} = \ee^{\langle PP\rangle_{0,0}} \ .
\end{equation}
To show that \eqref{constitutive} holds on the large phase space, it
is enough to show that the derivatives with respect to the descendant
couplings $t_{k,P}$ and $t_{k,H}$ vanish for $k\ge 1$. For this, let's
temporarily drop the $(0,0)$ subscript and consider
\begin{equation}
\eqlabel{enough}
\del_{t_{k,X}} \langle HH\rangle \overset{?}{=}
\del_{t_{k,X}} \ee^{\langle PP\rangle} \ ,
\end{equation}
where $X$ is $P$ or $H$, and we have turned on arbitrary values of
all the couplings. Now
\begin{equation}
\eqlabel{lhs}
\begin{split}
\del_{t_{k,X}} \langle HH \rangle &= \langle \sigma_k(X) HH\rangle \\
&=k \langle \sigma_{k-1}(X) H \rangle \langle P HH\rangle +
k\langle \sigma_{k-1}(X) P\rangle\langle HHH\rangle\\
&= k \langle \sigma_{k-1}(X) H\rangle \del_{t_{0,P}} \langle HH \rangle
+ k\langle\sigma_{k-1}(X) P\rangle\del_{t_{0,H}} \langle HH\rangle \ ,
\end{split}
\end{equation}
where we have used the topological recursion relations \cite{witten}.
On the other hand
\begin{equation}
\eqlabel{rhs}
\begin{split}
\del_{t_{k,X}} \ee^{\langle PP\rangle} &= \ee^{\langle PP\rangle}
\langle \sigma_k(X) PP \rangle \\
&= \ee^{\langle PP \rangle} \Bigl( k\langle\sigma_{k-1}(X) H\rangle
\langle PPP\rangle + k\langle\sigma_{k-1}(X) P\rangle
\langle HPP\rangle\Bigr) \\
&= k\langle\sigma_{k-1}(X) H\rangle \del_{t_{0,P}}\ee^{\langle PP\rangle}
+ k\langle\sigma_{k-1}(X) H \rangle \del_{t_{0,H}}\ee^{\langle PP\rangle} \ .
\end{split}
\end{equation}
Now \eqref{lhs} and \eqref{rhs} together with \eqref{constitutive} at
$t_{k,P}=t_{k,H}=0$ imply that the constitutive relations \eqref{constitutive}
hold on the large phase space as well.

\subsection{Adding boundaries}
In the open sector we have
the operator $B$  that fixes a point on the boundary of the worldsheet and
the primary operator $E$.
As we will argue, there are no gravitational descendants of the boundary primary operators.

Now let us add the A-brane wrapped on the equator of $\CP^1$, with trivial
gauge field, and consider the disk amplitude. Relevant instantons are now
maps
\begin{equation}
\eqlabel{diskmap}
(D,\del D)\to (\CP^1,\RP^1) \ .
\end{equation}
Any such map can be complex conjugated to a map from $\CP^1\to\CP^1$, and
we call the instanton number to be the ordinary degree of this doubled map.
Note however that the requirement that the boundary of the disk map to the
A-brane implies that the dimension of moduli space of such real maps is half
of what it was in the complex case.

Before coupling to topological
gravity, the selection rule is
\begin{equation}
\eqlabel{realcond}
1+2k = 2n_H+ m_E \ ,
\end{equation}
where $m_E$ is the number of insertions of the boundary operator $E$.
Note that \eqref{realcond} is an equality on {\it real dimensions}.
After coupling to gravity, the selection rule becomes
\begin{equation}
\eqlabel{except}
-2 + 2k + 2n_H + m_E + 2n_P + m_B = 2n_H + m_E \ .
\end{equation}

Now let us try to write down some correlators of primaries, after coupling to
topological gravity. First of all, the selection rule \eqref{except} allows
only solutions for $k=0$ and $k=1$. From $k=0$, we find a non-vanishing correlator
from $m_B=2$, $n_P=0$.
\begin{equation}
\langle BB E \rangle_{0,1} = 1 \ ,
\end{equation}
which simply comes from the constant map to the point dual to $E$. Similarly,
for $m_B=0$, $n_P=1$, we get
\begin{equation}
\langle P E \rangle_{0,1} =1 \ .
\end{equation}
However, any further insertion of $E$, or trying to insert $H$ instead of $E$,
gives a vanishing result, because a constant map can only map to a single point.
(Ultimately, this statement might require some rectification in view of \eqref{correct}
below.)

Now what about instanton sector 1, in which \eqref{except} implies we have no
insertions of puncture operators? First, note that there are two maps in this
sector: The map covering the northern half of the sphere, and the map covering
the southern half of the sphere. Second, consider correlators with only
boundary insertions. The first non-trivial one to consider is $\langle E^3\rangle$.
Before coupling to gravity, only one of the two possible instantons contributed to
this amplitude because of the fixing of the cyclic ordering of the boundary
insertions. Specifically, in the topological sigma model before coupling to
gravity, the amplitude is defined by
\begin{equation}
\eqlabel{definedby}
\langle E^3\rangle =\langle E(x_1) E(x_2) E(x_3)\rangle \ ,
\end{equation}
where the $x_i$ are some {\it fixed} insertion points on the boundary of the disk.
Thinking of the latter as the upper half plane with boundary the real line, we
can choose the three insertion points to be $x_1=0$, $x_1=1$, $x_3=\infty$.
Now to compute \eqref{definedby}, we choose three generic points $p_1$, $p_2$,
$p_3$ on the A-brane (equator of $\CP^1$), each representing the Poincar\'e dual of
the cohomology class $E$ generating $H^1(S^1,\zet)$, and count the number of
maps (instantons) mapping $x_i$ to $p_i$. It is not hard to see that depending
on the cyclic ordering of $p_1$, $p_2$, $p_3$, such a degree one map has to
cover either the northern or the southern hemisphere.

After coupling to topological gravity, the situation changes dramatically.
The definition of the amplitude now involves an {\it integration} over the
position of the insertion points, and dividing by the isometry group $SL(2,\reals)$
(in contrast, the definition of \eqref{definedby} a priori depends on $x_1$, $x_2$,
$x_3$). Although the $SL(2,\reals)$ symmetry of the disk still allows us to fix the
three insertion points at $0$, $1$, and $\infty$, we have to allow {\it both
possible cyclic orderings}. Hence, both hemispheres contribute to the correlator.
It us natural to assume that they contribute with opposite sign. (Justifying
this requires a more careful study of orientation of the relevant moduli spaces
from which we refrain here.) This implies that {\it after coupling
to topological gravity},
\begin{equation}
\eqlabel{vein}
\langle E^3\rangle_{0,1} = 0 \ .
\end{equation}
Continuing in this vein, we can add further boundary insertions. Each time,
both hemispheres contribute because we can always arrange both required
cyclic orderings of insertion points on the boundary of the disk. For
an even number of insertions, the two come with the same sign, and for
an odd number of insertions, we get a cancelation. (This is a consequence
of the fermionic nature of $E$.) Thus
\begin{equation}
\langle E^m\rangle_{0,1} = 1+(-1)^m \ .
\end{equation}
Finally, we consider correlators with insertion of $H$. A seemingly simple
 correlator to compute is $\langle EH\rangle$. Naively, one chooses a point
$p$ on the equator to represent $E$, and a point $q$ in the bulk to represent
$H$. Using $SL(2,\reals)$ invariance to put insertion of $H$ at $z=0$ the
center of the disk, and insertion of $E$ at say $x=1$ say, one looks for a
map that maps the center of the disk to $q$ and $x=1$ on the boundary to $p$.
There is a single such map in instanton sector $1$. It is the northern or
southern hemisphere depending on where $q$ is chosen. So one might conclude naively
\begin{equation}
\eqlabel{naively}
\langle EH\rangle_{0,1} = 1 \ .
\end{equation}
We claim, however, that this equation, {\it is wrong}. To see this, imagine we wanted to
compute $\langle E H^2\rangle_{0,1}$ by choosing a second bulk point $q'$ to
which we want to map a second bulk point $z'$. Since the map was already
fixed by $p$ and $q$, it boils down to the question of whether the second point
$z'$, in the interior of the disk (over which we are integrating) can be
chosen such that it maps to $q'$. The answer to this question {\it now depends
on whether we choose $q'$ in the northern or southern hemisphere}! But
clearly, the amplitude cannot depend on the choice of representative for
$H$, so something must be wrong with the reasoning leading to \eqref{naively}.

It is quite easy to see where the problem comes from on the worldsheet,
which also suggests the resolution. What we are trying to do is insert
\begin{equation}
\eqlabel{trying}
\int_D \Phi^* (\omega)
\end{equation}
into the path integral and claim that it gives something well-defined if
we think of $\omega$ as a cohomology class in $H^2(\CP^1)$. But clearly if
we change the representative $\omega\to\omega+d\psi$, \eqref{trying} changes
by a boundary term
\begin{equation}
\int_D \Phi^* (d\psi) = \int_{\del D} \Phi^*(\psi) \ .
\end{equation}
The well-known way to resolve this is to add explicit boundary term
in the form of the Wilson line
\begin{equation}
\int_{\del D} A \ .
\end{equation}
Mathematically, speaking, we have to think of $\omega$ as a relative cohomology class
in $H^2(\CP^1,\RP^1)$. Thinking this through, we learn that the correct and invariant
way to represent the operator $H$ is as the sum of a point in one of the
hemispheres of $\CP^1$ together with $1/2$ times a point on the equator. Specifically,
we need a point $p_0$ on the equator, oriented such that its intersection with
the equator in positive (east-west) direction is $+1$, and in negative direction
$-1$. Then if $q_N$ is a generic point in the northern hemisphere, we can
represent $H$ as a relative cohomology class by
\begin{equation}
\eqlabel{correct}
H \leftrightarrow q_N - \frac 12 p_0
\end{equation}
(alternatively, we could use a point $q_S$ in the southern hemisphere to
represent $H$ as $q_S+\frac 12 p_0$). How does this identification
repair \eqref{naively} and the conundrum below it? Well, the net effect will
be that both hemispheres contribute to the amplitude, canceling each
other, so that $\langle E H \rangle_{0,1}=0$.

But to justify this, it is easier to consider an amplitude with more insertions
so that we have an actual moduli space to play with. Consider for example
adding a bulk insertion to \eqref{vein}, in other words the amplitude
$\langle E^3 H\rangle$. Having represented $H$ by $q_N-\frac 12 p_0$ means that
the map to the northern hemisphere contributes $+1$, when the bulk insertion is
at a regular point of the disk. But in the compactification of the moduli
space of the disk with marked points, we have to bubble off disks whenever
bulk insertions approach the boundary. With three marked points on the boundary,
the added configuration consists of two disks, joined at a common node. The
first disk carries four marked boundary points (the three original ones plus the
node), and the other the bulk point plus the node on the boundary. In mapping
this to $(\CP^1,\RP^1)$, we collapse the bubbled disk. Since after the collapse,
we are just required to map the node to $p_0$, which is a point on the
equator, we can do this with either the northern or southern hemisphere. They
contribute with the same sign. Thus, the northern hemisphere contributes
$1-\frac 12=1/2$, while the southern hemisphere contributes $-\frac 12$,
for a total amplitude of
\begin{equation}
\langle E^3 H\rangle_{0,0} = \frac 12 - \frac 12 = 0 \ .
\end{equation}

We can extract the general logic from this example: An extra insertion of $H$,
represented rationally by $q_N-\frac 12 p_0$ changes the contribution of the northern
hemisphere by $1-\frac 12$, where the $1$ comes from a bulk insertion being mapped
to $q_N$, and the $-\frac 12$ comes from the bulk insertion moving to the boundary,
bubbling off a disk, which is subsequently collapsed to an additional boundary
insertion, and mapped to $p_0$. The contribution of the southern hemisphere changes
by a factor of $\frac 12$, coming entirely from the bubbled configuration.
(The sign is negative of the bubbled configuration on the northern hemisphere
because the boundary is oriented oppositely.) With very few insertions, the argument
is somewhat delicate to carry out, but the general conclusion is
\begin{equation}
\langle H^n E^m \rangle_{0,0} = \frac 1{2^n} \bigl(1+(-1)^m\bigr)
\qquad 2n+m\ge 3 \ .
\end{equation}
These results can be summarized in the generating function
\begin{equation}
\eqlabel{superpotential}
F^{(0,1)} = \frac 12 t_{0,B}^2 t_{0,E} + t_{0,P} t_{0,E}
+ \ee^{t_{0,H}/2} \bigl( \ee^{t_{0,E}} + \ee^{-t_{0,E}}\bigr) \ ,
\end{equation}
where the polynomial piece is somewhat unclear at this point, for reasons
explained above.

We are now prepared to turn on gravitational descendants and study the
analogue of the constitutive relations.

\subsection{Constitutive relations with D-branes}

First of all, let us argue that there are no gravitational descendants of boundary operators.
If there were any, they should have a cohomological definition,
most likely involving the tangent space to the boundary, fitting together
to a real line bundle over the moduli space $\calm_{g,h}$. But the only characteristic
class of such a real line is the first Stieffel-Whitney class, which is
a torsion class in $H^1(\zet_2)$. Since one cannot build intersection
theory on torsion classes, it is pretty much excluded that one can have
gravitational descendants as actual ``local operators''. On the other
hand, and guided by intuition gained from mirror symmetry on threefolds, see
\eg, \cite{opening}, it seems likely that the torsion classes
can actually be used to define discrete observables similar to domain walls. In other
words, we have not derivatives of amplitudes with respect to would-be couplings, but finite
differences between certain ``D-brane vacua''.

Second, we make the assumption that the primary operators on the boundary
should be on-shell. Namely such that all one-point functions $\langle E \cdots
\rangle$, $\langle B \cdots \rangle$, where $\cdots$ are arbitrary bulk insertions,
should vanish. Referring back to \eqref{superpotential}, we learn that $t_{0,B}=0$,
and $\ee^{t_{0,E}}= \pm 1$. These two solutions are precisely the two choices
of Wilson lines that we have met before.
This decision to freeze open string moduli is again motivated by the Calabi-Yau
threefold case. We will see that it is a very useful technical assumption that
allows solving the theory completely (which we will do at tree-level).

Indeed, without boundary couplings, the small and large phase space are unchanged
from the purely closed string case. We just have additional observables. So we begin
by writing the disk amplitudes as functions of the topological coordinates
\eqref{coordinates}. Consulting \eqref{superpotential}, we see that on-shell,
\begin{equation}
\eqlabel{squareroot}
\langle H\rangle_{0,1} = \pm \ee^{t_{0,H}/2} = \pm \ee^{\langle PP\rangle_{0,0}/2} \ .
\end{equation}
This equation can be viewed as a natural squareroot of \eqref{constitutive},
very much in agreement with the ``real topological string paradigm'' developed
in \cite{tadpole,realts}

We now claim that \eqref{squareroot} also holds on the large phase space, with
non-zero (bulk) descendant couplings $t_{k,H}, t_{k,P}$. To show this, we proceed as
in \eqref{enough}. The derivative of the right hand side is
\begin{equation}
\eqlabel{agreement}
\del_{t_{k,X}} \ee^{\langle PP\rangle_{0,0}/2}
=k \langle \sigma_{k-1}(X) H\rangle_{0,0} \del_{t_{0,P}}\ee^{\langle PP\rangle_{0,0}/2}
+ k\langle\sigma_{k-1}(X)P\rangle_{0,0}\del_{t_{0,H}}\ee^{\langle PP\rangle_{0,0}/2} \ .
\end{equation}
Let us check that the left-hand side gives the same
\begin{equation}
\del_{t_{k,X}} \langle H\rangle_{0,1} = \langle \sigma_{k}(X) H\rangle_{0,1} \ .
\end{equation}
We wish to rewrite this using topological recursion relations. There are
two possible degenerations that could make a contribution. The first is when the
two bulk insertions come close together, leading to the bubbling of a sphere at
the center of the disk. The local situation is as on the sphere, so we can readily
copy \cite{witten} to conclude that the contribution is
\begin{equation}
\eqlabel{only}
k \langle\sigma_{k-1}(X) H\rangle_{0,0} \langle PH\rangle_{0,1}
+ k\langle\sigma_{k-1}(X)\rangle_{0,0} \langle HH\rangle_{0,1}
\end{equation}
The second degeneration occurs when a bulk insertion moves to the boundary,
leading to the bubbling of a disk, as we have seen in our discussion of
the off-shell correlators \eqref{superpotential}. On-shell however, such
degenerations make no contribution because disk one-point functions vanish.
Thus we find
\begin{equation}
\del_{t_{k,X}}\langle H\rangle_{0,1} =
k\langle\sigma_{k-1}(X)H\rangle_{0,0}\del_{t_{0,P}}\langle H\rangle_{0,1}
+ k\langle\sigma_{k-1}(X)P\rangle_{0,0}\del_{t{0,H}}\langle H\rangle_{0,1}
\end{equation}
and comparison with \eqref{agreement} shows as in the bulk that \eqref{squareroot}
holds on the large phase as well.

\section{Discussion}

In the paper, we made some steps towards an understanding of the summation over
boundaries in topological field theories and in topological strings.
We calculated the correlation functions of the bulk and boundary operators in
topological sigma-model on $\CP^1\cong S^2$,
studied the symmetries of the model
and the open/closed TFT correspondence.
We then coupled the model to topological gravity and derived constitutive
relations between the correlation functions of bulk and boundary operators.

There are various directions that are worth exploring, in particular when coupling to topological
gravity. Already before summing over boundaries it is
interesting to ask how does the coupling to topological gravity affect the duality relation in
the closed TFT. In the case of $\CP^1$ target space, the partition function is related to the tau
function of the Toda hierarchy (see e.g. \cite{Aganagic:2003qj}), and one may consider the symmetries of the latter.

Integrability is an important issue when considering topological gravity already without matter
TFT but in the presence of boundaries. When there are no boundaries we have the generating function
for the correlators
\beq
F(t_0,t_1,...) = \langle \exp \Big(\sum_{i=1}^{\infty} t_i \sigma_i \Big) \rangle
\eeq
is the $\tau$-function of the KdV hierarchy \cite{witten}. Is there an analogous structure with the
inclusion of boundaries and the boundary puncture operator?

An analysis of the coupling of the model to topological gravity seems most rewarding. In particular,
can one justify more rigorously our statements about open topological gravity such as the absence of
boundary descendants. It is also a reasonable question to ask for the differential equations
satisfied by the all-genus, all-boundaries partition function.

There is a relation between the partition function of the closed A-model on  $\CP^1$
and the generating functional for certain correlation functions of supersymmetric gauge theory on $R^4$ \cite{Losev:2003py,Nekrasov:2009zza}.
It would be natural to ask whether there is an analogous relation in the presence of boundaries.

Finally, it would be interesting to generalize the summation over the boundaries in to other cases.
One may expect, for instance,  a straightforward generalization to $\CP^n$ Sigma Models \cite{Melnikov:2005tk}.

\begin{acknowledgments}
This project was started a very long time ago, and we acknowledge the hospitality of several
institutions, especially, KITP Santa Barbara, IAS Princeton, Hebrew University, and CERN, over
the course of the collaboration. E.R.\ would like to thank Nikita Nekrasov and Don Zagier
for valuable discussions related to this work.
The work is supported in part by the Israeli
Science Foundation center of excellence, by the Deutsch-Israelische
Projektkooperation (DIP), by the US-Israel Binational Science
Foundation (BSF), and by the German-Israeli Foundation (GIF).
\end{acknowledgments}

\end{document}